\documentclass[reprint,superscriptaddress,nofootinbib,amsmath,amssymb,aps,pra,floatfix]{revtex4-1}
\usepackage{caption}
\usepackage{graphicx}
\usepackage{dcolumn}
\usepackage[inline]{enumitem}
\usepackage{hyperref}
\usepackage{pifont}

\usepackage{bm}
\usepackage{amsmath}
\usepackage[font=small]{caption}
\usepackage[labelformat=empty, position=top]{subcaption}
\usepackage[export]{adjustbox}
\usepackage{blindtext}
\usepackage{amssymb}
\usepackage{newunicodechar}
\usepackage{nameref}
\usepackage{xcolor}

\begin{document}

\title{Graph Neural Network-based EEG Classification: A Survey}

\author{Dominik Klepl}
  \affiliation{Centre for Computational Science and Mathematical Modelling, Coventry University, Coventry CV1 2JH, UK 
  }

\author{Min Wu}
  \affiliation{Institute for Infocomm Research, Agency for Science, Technology and Research (A*STAR), 138632, Singapore
}

\author{Fei He}
  \email{Correspondence to: fei.he@coventry.ac.uk}
  \affiliation{Centre for Computational Science and Mathematical Modelling, Coventry University, Coventry CV1 2JH, UK
}


\begin{abstract}
Graph neural networks (GNN) are increasingly used to classify EEG for tasks such as emotion recognition, motor imagery and neurological diseases and disorders. A wide range of methods have been proposed to design GNN-based classifiers. Therefore, there is a need for a systematic review and categorisation of these approaches. We exhaustively search the published literature on this topic and derive several categories for comparison. These categories highlight the similarities and differences among the methods. The results suggest a prevalence of spectral graph convolutional layers over spatial. Additionally, we identify standard forms of node features, with the most popular being the raw EEG signal and differential entropy. Our results summarise the emerging trends in GNN-based approaches for EEG classification. Finally, we discuss several promising research directions, such as exploring the potential of transfer learning methods and appropriate modelling of cross-frequency interactions.

\end{abstract}

\maketitle

\section{Introduction}
Electroencephalography (EEG) is a non-invasive technique used for recording electrical brain activity with a wide range of applications in cognitive neuroscience \cite{morales_time-frequency_2022}, clinical diagnosis \cite{smith_eeg_2005, loo_clinical_2005}, and brain-computer interfaces \cite{mcfarland_eeg-based_2017, lotte_review_2018}. However, analysing EEG signals poses several challenges, including a low signal-to-noise ratio, non-stationarity resulting from brain dynamics, and the multivariate nature of the signals \cite{samek2012stationary, klepl2022GNN}. In this review, we focus on the classification of EEG, such as emotion recognition, motor imagery recognition or neurological disorders and diseases. 

Traditional feature extraction methods for EEG classification, such as common spatial patterns \cite{samek2012stationary}, wavelet transform \cite{murugappan2010wavelet}, and Hilbert-Huang transform \cite{oweis2011hilbert_huang}, have been commonly employed. These methods aim to extract meaningful features from EEG signals \cite{hosseini_review_2021, rasheed_machine_2021}, with key features like power spectral density (PSD) \cite{klepl2022GNN} to characterise brain states. However, relying on such manually defined features to train machine learning classifiers has several limitations. Subjectivity and biases in feature selection, along with time-consuming engineering and selection processes, limit scalability and generalisation \cite{klepl2023bispectrum, klepl2022GNN}. Automated feature extraction methods are needed to overcome these limitations, improve efficiency, reduce bias, and enhance classifier adaptability to different EEG datasets.

Deep learning architectures, such as convolutional neural networks (CNN) and long short-term memory (LSTM) networks, have also been explored for EEG analysis \cite{lawhern2018eegnet, hefron2017LSTM}. However, they face challenges in effectively capturing the spatial dependencies between electrodes and handling the temporal dynamics of EEG signals \cite{klepl2022GNN}. Modelling the complex sequential and spatial relationships in EEG data is crucial for more accurate classification and analysis.

Network neuroscience offers an alternative approach to EEG modelling by framing the signals as a graph. The brain exhibits a complex network structure, with neurons forming connections and communicating with each other \cite{bassett2017network}. Analysing EEG data as a graph enables the study of network properties, including functional connectivity, providing insights into brain function and dysfunction \cite{klepl2023bispectrum, cao2020FC, adebisi_brain_2023}. Graph-based analysis facilitates the examination of network features, node importance, community structure, and information flow, offering insights into brain organisation and dynamics. Such graph-theory-based features were shown to be powerful predictive features for EEG classification \cite{klepl2023bispectrum, kilicc2022graphtheory, jalili2017graphtheory, hasanzadeh2020graphtheory, nobukawa2020graphtheory, adebisi_brain_2023, supriya_epilepsy_2023}. However, these features have the same limitations as manually defined features based on traditional EEG analysis methods introduced above.

Graph Neural Networks (GNNs) emerge as a powerful tool for modelling neurophysiological data \cite{li_graph_2023}, such as EEG, within the network neuroscience framework \cite{klepl2022GNN, klepl2023adaptive}. GNNs are specifically designed to operate on graph-structured data. They can effectively leverage the spatial structure within EEG data to extract features, uncover patterns and make predictions based on the complex interactions between different electrodes. Designing GNN models for EEG classification will likely improve classification tasks and potentially uncover new insights in neuroscience.

\begin{figure*}[ht]
    \centering
    \includegraphics[width = 0.95\textwidth]{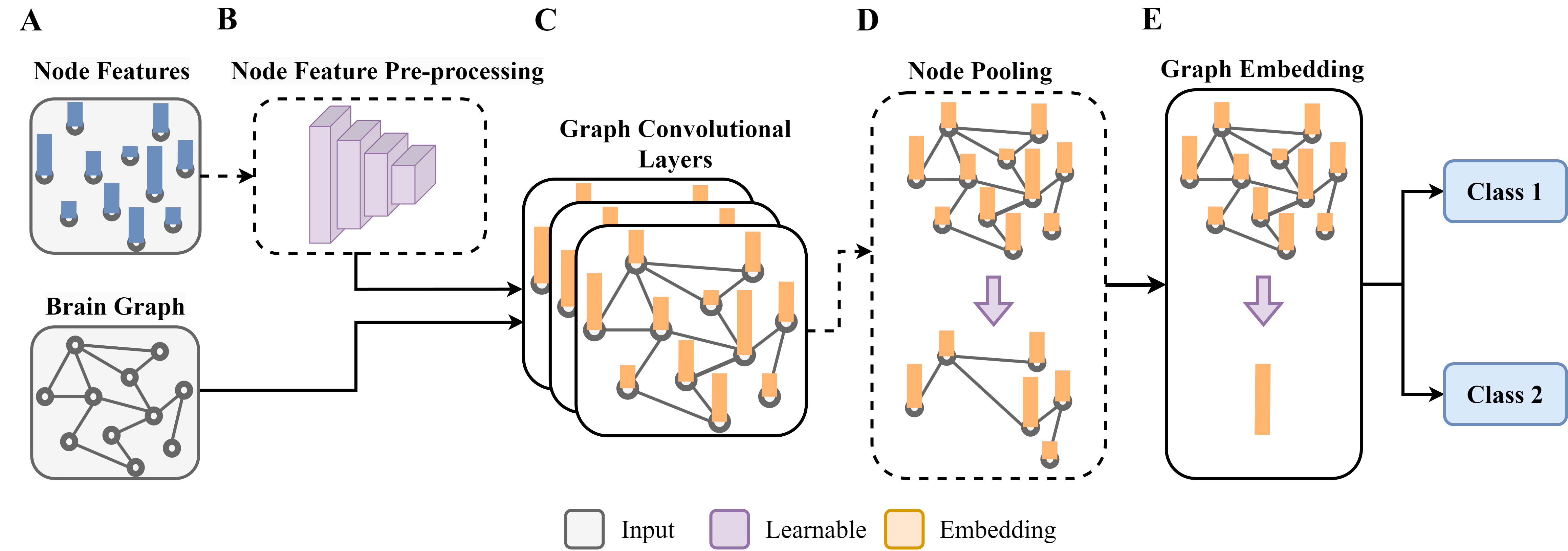}
    \caption{General architecture of a graph neural network model for classification of EEG. (A) The input to the model consists of node features and a possibly learnable brain graph structure. (B) Optionally, the node features can undergo pre-processing via a neural network. (C) Next, the node features are passed to a block of graph convolutional layers, where node embeddings are learned. (D) Then, a node pooling module can be utilised to coarsen the graph. Node pooling may contain learnable parameters as well. (E) Finally, the set of node embeddings forms a graph embedding, which can be used to predict the outcome.}
    \label{fig:overview}
\end{figure*}

Motivated by the potential of GNNs and an increasing number of recent papers proposing GNN for various EEG classification tasks, there is an urgent need for a comprehensive review of GNN models for EEG classification. The main contributions of this paper include:

\begin{itemize}
\item Identifying emerging trends of GNN models tailored for EEG classification.
\item Reviewing popular graph convolutional layers and their applicability to EEG data.
\item Providing a unified overview of node feature and brain graph structure definitions in the context of EEG analysis.
\item Examining techniques for transforming sets of node feature embeddings into a single graph embedding for graph classification tasks.
\end{itemize}

By addressing these essential aspects, this review paper will provide a comprehensive and in-depth analysis of the application of Graph Neural Network (GNN) models for EEG classification. The findings and insights gained from this review will serve as a resource to navigate this emerging field and identify promising future research directions.

\section{Overview of Graph Neural Networks}
\label{section: background}
Graphs are widely used to capture complex relationships and dependencies in various domains, such as social networks, biological networks, and knowledge graphs. The problem of graph classification, which aims to assign a label to an entire graph, has gained attention in recent years. GNNs offer a promising solution to this problem by extending the concept of convolution from Euclidean inputs to graph-structured data. GNNs have been successfully applied in a wide range of fields, such as biology \cite{li_graph_2023}, bioinformatics \cite{zhang_graph_2021}, network neuroscience \cite{bessadok_graph_2023}, chemistry \cite{wieder_compact_2020, reiser_graph_2022}, drug design and discovery \cite{xiong_graph_2021, sun_graph_2020}, natural language processing \cite{malekzadeh_review_2021, wu_graph_2023}, recommendation systems \cite{gao_survey_2023, wu_graph_2022}, traffic prediction \cite{jiang_graph_2022, lv_temporal_2021} and finance \cite{wang_review_2022}.

In graph classification problems, the input is a set of graphs, each with its own set of nodes, edges, and node features. Let $G = (V, E, H)$ denote a featured graph, where $V$ represents the set of nodes, $E$ represents the set of edges connecting the nodes, and $H$ represents the $V \times D$ matrix of $D$-dimensional node features. In the case of EEG, the EEG channels are the nodes, and edges represent structural or functional connectivity between pairs of nodes. Each graph $G$ is associated with a label $y$, indicating its class. The goal is to learn a function $f(G) \rightarrow y$ that can predict the class label $y$ given an input graph $G$. A general structure of a GNN model for EEG classification is presented in Fig \ref{fig:overview}.

Compared to other deep learning models, GNNs offer several advantages. First, GNNs were specifically designed for graph-structured inputs. This means that GNNs can adapt to irregularly structured inputs, i.e. graphs with varying numbers of nodes, compared to traditional deep learning, such as CNN, that require fixed-size inputs. Next, GNNs can simultaneously learn information from node features and the graph structure by accepting two inputs: node feature matrix and graph structure. Such simultaneous integration is not possible with traditional deep learning methods.

\begin{figure*}[tb]
    \centering
    \includegraphics[width = 0.95\textwidth]{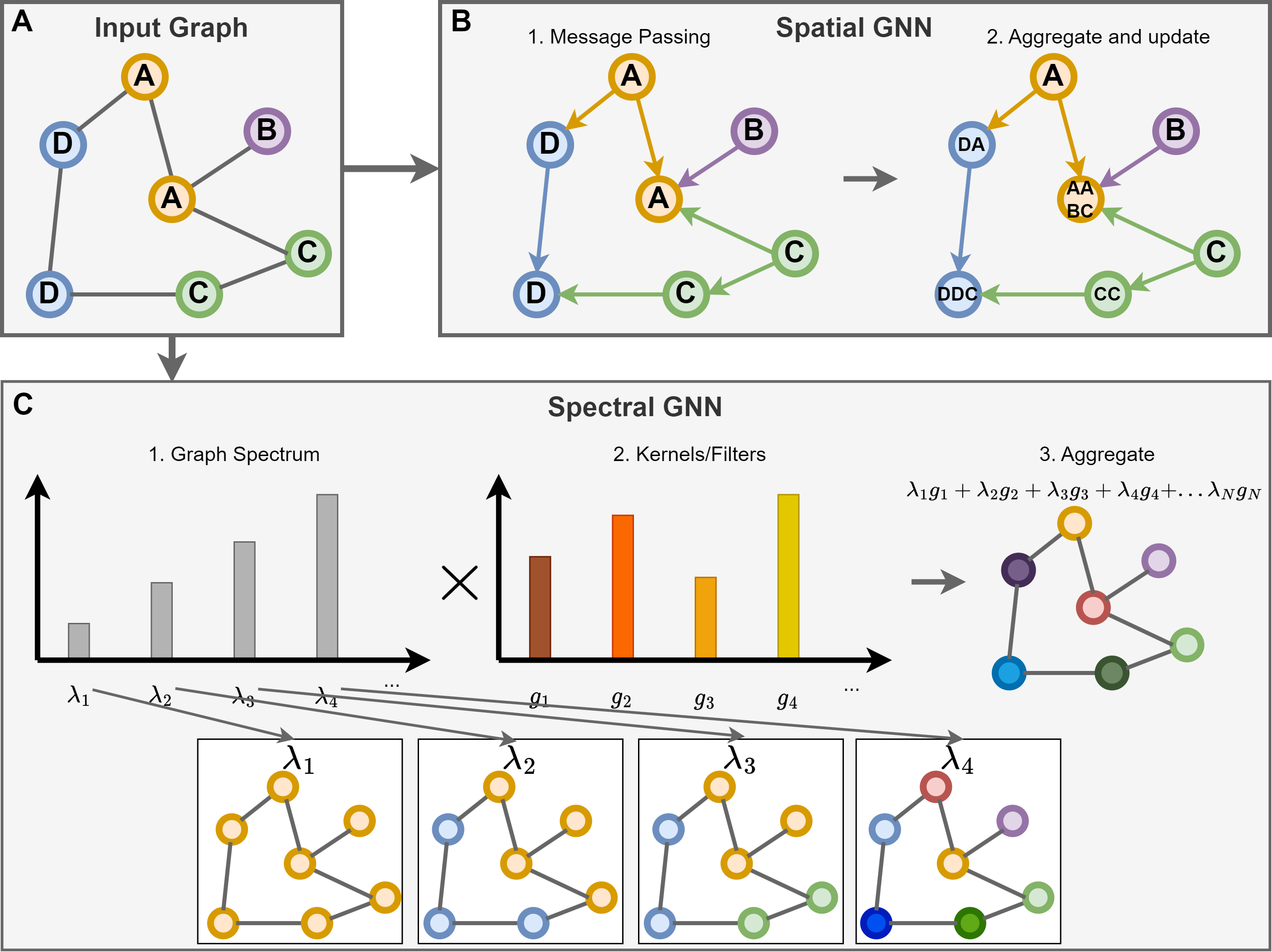}
    \caption{Illustration of core mechanisms of spatial and spectral GNNs. A) An undirected featured graph is given as an example input graph with node features shown as node labels and colours. B) Spatial GNNs operate in the graph domain directly using message passing to update node embeddings. 1) Messages, i.e. transformed node features or embeddings, are sent along edges. For simplicity, we show only one direction of the flow of messages. 2) The collected messages at each node are aggregated using a permutation-invariant function and are fused with the original node embedding to form an updated node embedding. Thus, one spatial GNN layer results in node embeddings containing information about the 1-hop neighbourhood of a given node. Thus, $L$ layers are required for node embeddings to access the information from the $L$-hop neighbourhood. C) In contrast, spectral GNNs operate in the graph spectral domain. 1) Node features are treated as signals on top of a graph and are deconstructed into graph frequencies given by the eigendecomposition of the graph Laplacian. Graph frequencies can be interpreted as variations of the signals. 2) The contribution of each graph frequency is weighted by the set of learnable kernels $G$ that effectively function as graph filters. 3) Node embeddings are then obtained by aggregating the filtered graph frequencies and transforming them back to the spatial graph domain. Thus, full spectral GNNs can access information from $N$-hop neighbourhoods where $N$ is the number of nodes of a given graph. However, in practice, approximations such as Chebyshev graph convolution restrict this to the chosen hop size.}
    \label{fig:convolution}
\end{figure*}

Multiple types of GNNs have been well introduced in \cite{wu_comprehensive_2021, zhou_graph_2020}. In this survey, we briefly introduce the two main branches of GNNs, namely, spatial and spectral GNNs (Fig. \ref{fig:convolution}). Other types of GNNs, such as attention GNNs \cite{velickovic_gat_2018}, recurrent GNNs \cite{seo_structured_2018}, and graph transformers \cite{shi_masked_2021}, can be viewed as special cases of spatial GNNs, and thus we will not provide detailed discussion in this survey. Both spatial and spectral GNNs aim to extend the convolution mechanism to graph data. For a detailed review of their similarities and differences, see \cite{chen_bridging_2021}. Moreover, for a comparison of GNNs in terms of computational complexity, see \cite{wu_comprehensive_2021}.

Spatial GNNs aggregate information from neighbouring nodes, similar to traditional convolution applied to image data aggregating information from adjacent pixels. Stacking multiple spatial GNN layers leads to information aggregation from various scales going from local to global patterns being captured in early and later layers, respectively. In contrast, spectral GNNs perform information aggregation in the graph frequency domain, with low-frequency and high-frequency components capturing global and local patterns, respectively. However, both approaches learn to capture local and global patterns within the graph, i.e. high and low-frequency information in the spectral domain. The advantage of spectral GNNs is their connection to graph signal processing, allowing for interpretation from the perspective of graph filters. However, spectral GNNs do not generalise well to large graphs since they depend on the eigendecomposition of graph Laplacian. In contrast, spatial GNNs can be applied to large graphs since they perform only local message-passing. On the other hand, spatial GNNs may be challenging to interpret and prone to overfitting because of over-smoothing, where embeddings of all nodes become similar.

\subsection{Spatial GNNs}

Spatial GNNs directly operate on the graph structure via the adjacency matrix operator. Given a set of nodes and associated features, spatial GNNs perform neighbourhood aggregation to derive node embeddings. This process is referred to as message passing. Intuitively, nodes connected by edges should have similar node embeddings, i.e. local node similarity. Message passing implements this idea by updating node embeddings with aggregated information collected from the node's neighbourhood. Formally, the node update equation in $l$\textsuperscript{th} layer of spatial GNN with $L$ layers is defined as follows:

\begin{equation}
h_i^{(l+1)} = \sigma\left(W_1^{(l)}h_i^{(l)} + \sum_{j \in \mathcal{N}(v_i)} W_2^{(l)}h_j^{(l)}e_{ji} \right),
\label{eq: spatial GNN}
\end{equation}

where $h_i$ is the node embedding vector, or when $l=1$, this is the input node feature vector, $\sigma$ is the activation function, $\sum$ is the aggregation function, $\mathcal{N}(v_i)$ is the neighbourhood of node $v_i$, $W \in \mathbb{R}^{d_1 \times d_2}$ is a learnable parameter matrix projecting node embeddings from input dimension $d_1$ to hidden dimension $d_2$ and $e_{ji}$ is the edge weight ($e_{ji}=1$ for unweighted graphs).

A single spatial GNN layer aggregates information from the $1$-hop neighbourhood. Thus, to increase the reception field of the model, $L$ spatial GNN layers can be stacked to aggregate information from up to $L$-hop neighbourhoods. A disadvantage of spatial GNNs is the difficulty of training deep models with many layers. With an increasing number of layers, the node embeddings become increasingly smooth, i.e. variance among embeddings of all nodes decreases. This happens when the messages already contain aggregated information from the whole graph; continual message passing of such saturated messages leads to oversmoothing, i.e., all node embeddings becoming essentially identical.

\subsection{Spectral GNNs}
Spectral GNNs can also be applied to EEG classification tasks by leveraging the spectral domain analysis of graph-structured data. 
The EEG graph is transformed into the spectral domain using the Graph Fourier Transform (GFT) and Graph Signal Processing (GSP) techniques. For a detailed review of spectral GNN methods, please refer to \cite{bo_2023spectralGNN_survey}.

The graph spectrum is defined as the eigendecomposition of the graph Laplacian matrix. The GFT is then defined as $\mathbf{\hat{H}} = \mathbf{U}^T \mathbf{H}$, its inverse as $\mathbf{H} = \mathbf{U} \mathbf{\hat{H}}$, where $\mathbf{U}$ is the orthonormal matrix of eigenvectors of the graph Laplacian $\mathbf{L}$ and $H \in \mathbf{R}^{N \times D}$ is the matrix of node feature vectors with $N$ and $D$ being the number of nodes and dimensionality of node features, respectively. The graph Laplacian is defined as $\mathbf{L} = \mathbf{D}-\mathbf{A}$, but often the normalised version is preferred: $\mathbf{\hat{L}} = \mathbf{I} - \mathbf{D}^{-1/2} \mathbf{A} \mathbf{D}^{-1/2}$ ($\mathbf{A}$ and $\mathbf{D}$ are the adjacency and degree matrices, respectively).

Spectral GNN is then typically defined as the convolution ($*$) of a signal defined on graph $\mathbf{H}$ and a spatial kernel $g$ in the spectral domain, thus becoming an element-wise multiplication ($\odot$):

\begin{equation}
    \mathbf{H} * g = \mathbf{U} \left( \left( \mathbf{U}^T \mathbf{H} \right) \odot \left( \mathbf{U}^T g \right) \right).
\end{equation}

Generally, $\mathbf{U}^T g$ is defined as a learnable diagonal matrix $\mathbf{G} = diag(g_1,...,g_V)$ spectral filter \cite{bo_2023spectralGNN_survey}.

However, the full spectral graph convolution can be computationally expensive. A popular approximation is the Chebyshev GNN (ChebConv) \cite{defferrard2016ChebGCN}, which performs localised spectral filtering on the graph. The node embedding update equation of a ChebConv is defined as:

\begin{equation}
H * g \approx \sum^{K}_{i=1} \mathbf{\Theta}_i T_i(\mathbf{\hat{L'}}),
\end{equation}

where $\mathbf{\Theta} \in \mathbb{R}^{K \times d \times d}$ are learnable parameters, $T_i(\mathbf{\hat{L'}}) = 2T_{i-1}(\mathbf{\hat{L'}}) - T_{i-2}(\mathbf{\hat{L'}})$, $T_1(\mathbf{\hat{L'}}) = \mathbf{H}$, $T_2(\mathbf{\hat{L'}}) = \mathbf{\hat{L'}} \mathbf{H}$, and $\mathbf{\hat{L'}} = \frac{2\mathbf{\hat{L}}}{\lambda_{max}} - \mathbf{I}$ ($\lambda_{max}$ is the largest eigenvalue of $\mathbf{\hat{L}}$, often approximated as $\lambda_{max} = 2$). The $K$ parameter controls the size of the Chebyshev filter.

However, spectral GNNs are limited to input graphs with a fixed number of nodes. This is because of the explicit use of the graph Laplacian. This is in contrast to spatial GNNs, which do not rely on explicitly materialising the adjacency matrix.

\section{Survey results}
\begin{figure}[ht]
    \centering
    \includegraphics[width = 0.98\linewidth]{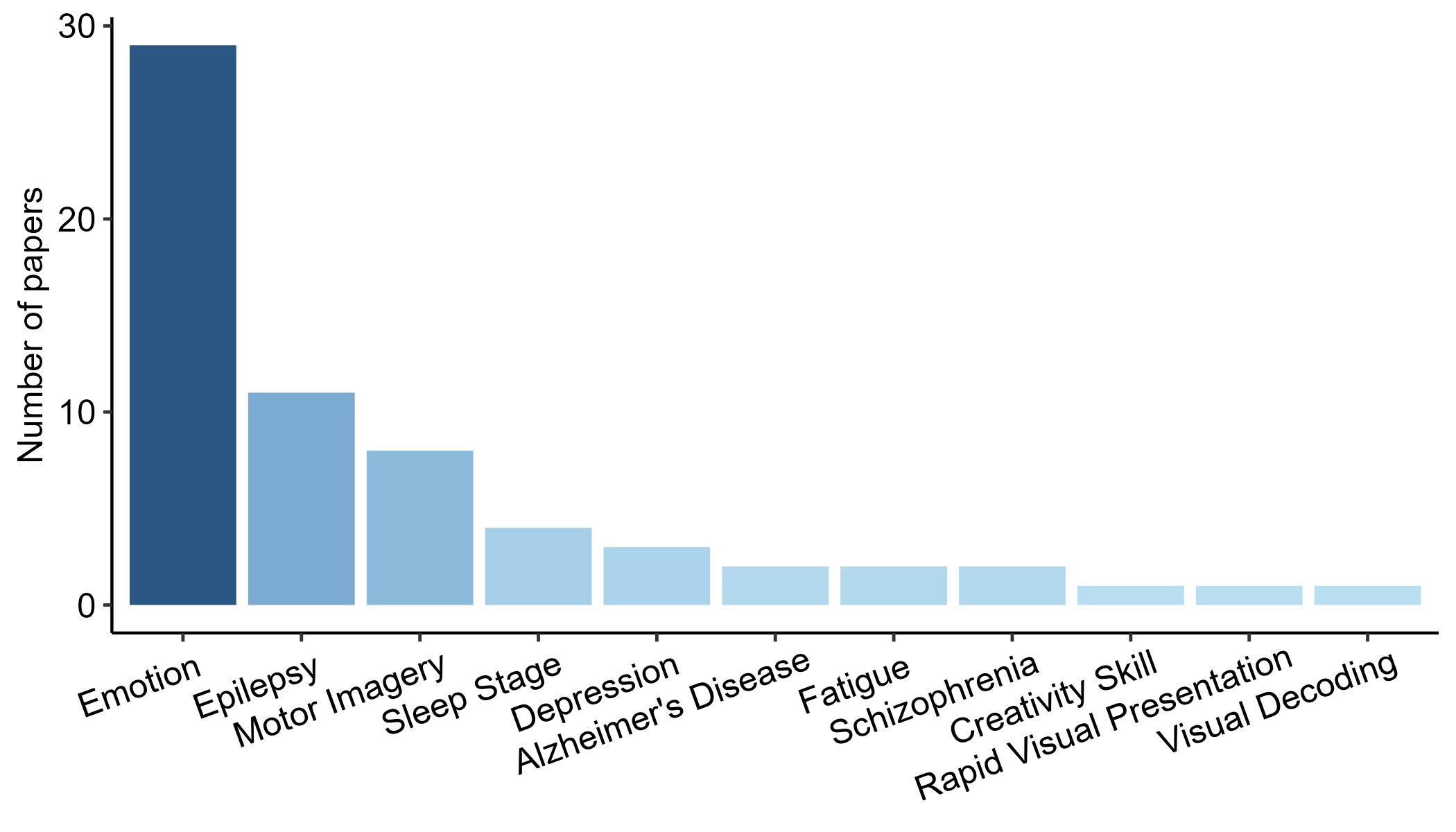}
    \caption{Classification tasks presented in the current EEG-GNN literature.}
    \label{fig: classification task}
\end{figure}

This survey is based on a review of 63 articles. These articles were selected by title and abstract screening from a search on Google Scholar and ScienceDirect queried on November 1st, 2022. The search query for collecting the articles was defined as: (``Graph neural network" OR ``Graph convolutional network") AND (``Electroencephalography" OR "EEG"). Both peer-reviewed articles and preprints were searched and utilised. All types of EEG classification tasks were included. We summarise the various types of EEG classification tasks identified in the surveyed papers in Fig \ref{fig: classification task}. The most common classification tasks are emotion recognition, epilepsy diagnosis and detection and motor imagery. However, the type of classification task should have a relatively minor effect on the GNN architecture design. Thus, we do not analyse and discuss this in detail. Instead, we survey the various GNN-based methods for EEG classification, intending to systematically categorise the types of GNN modules and identify emerging trends in this field independent of the specific classification task.

In the remaining portion of this paper, we report the categories of comparisons we identified in the surveyed papers. These are based on the different modules of the proposed GNN-based models. Specifically, these are:

\begin{itemize}
  \item Definition of brain graph structure
  \item Type of node features
  \item Type of graph convolutional layer
  \item Node feature preprocessing
  \item Node pooling mechanisms
  \item Formation of graph embedding from the set of node embeddings
\end{itemize}

The following sections will provide further details on these categories, and the paper will conclude by discussing trends and proposing plausible directions for future research.

\section{Definition of Brain graph structure}
\begin{table*}[ht]
\centering
\caption{Overview of methods for obtaining the brain graph structure.} 
\begin{tabular}{|m{0.25\textwidth}<{\centering}|m{0.08\textwidth}<{\centering}|m{0.09\textwidth}<{\centering}|m{0.48\textwidth}<{\centering}|}
  \hline
Method & Learnable & Pre-defined & Papers \\ 
  \hline
Distance between electrode positions & \ding{55} & \ding{51} & \cite{zhong2020GNN-regularised-emotion, lin2021driving, demir2021ERP-comparison, raeisi2022neonatal-epilepsy, tang2021selfsupervised, yin2021LSTMfusion, sun2021adaptive, du2022multi, xu2022dagam, jang2018video, ding2021lggnet, zheng2021hiearachy, zeng2020hierarchy, jia2022cr-gcn, priyasad2022affect, jia2022efficient, chen2022exploring, li2022attentionsleep, jia2021multi} \\ 
  Functional connectivity measure & \ding{55} & \ding{51} & \cite{liu2022MST-GNN, lin2021driving, kwak2020BCI-Fusion-GNN, chang2021MMN-schizophrenia, bi2022multi, raeisi2022neonatal-epilepsy, asadzadeh2022bayes, zhao2021seizure-focalloss, tang2021selfsupervised, chen2021bayesianGNN, li2021multidomain, ghosh2021decoding, khaleghi2023visualdecoding, wang2021depression, hou2020biLSTM-GCN, hou2022gcns, jang2018video, kong2022causalGNN, jia2022cr-gcn, tao2022seizure, wang2022coteaching, li2021mutualgraphnet, bao2022linking, feng2022eeg, he2022spatial, jia2022efficient, sun2022complex, klepl2022GNN, shan2022STGCN-AD} \\ 
  Manually defined & \ding{55} & \ding{51} & \cite{zhong2020GNN-regularised-emotion, li2022eeg, du2022multi, xu2022dagam, zheng2021hiearachy, zeng2020hierarchy, chen2022exploring} \\ 
  Shared learnable mask & \ding{51} & \ding{55} & \cite{zhong2020GNN-regularised-emotion, li2022attention-dynamic, asadzadeh2022bayes, li2021multidomain, sun2022transformer, zhang2019gcb, li2021active, xue2022adaptive, bao2022linking, li2022attentionsleep} \\ 
  Feature similarity & \ding{51} & \ding{55} & \cite{li2021self-organising-emotion, li2022eeg, li2021multidomain, ding2021lggnet, xue2022adaptive, li2022attentionsleep, lu2022pearnet} \\ 
  Feature distance & \ding{51} & \ding{55} & \cite{jia2020graphsleepnet, wiercinski2022emotion, zeng2022siam, jia2021multi} \\ 
  Transformer-style attention & \ding{51} & \ding{55} & \cite{zhu2022locally, li2021attentionLSTM} \\ 
  Concatenation attention & \ding{51} & \ding{55} & \cite{li2021dualGAT, li2021mutualgraphnet} \\ 
  Dense projection & \ding{51} & \ding{55} & \cite{song2020instance, song2021variationalinstance, lian2020learning} \\ 
  LSTM-based & \ding{51} & \ding{55} & \cite{dissanayake2021geometric} \\ 
  Multiple/Combined graph definitions & \textbf{-} & \textbf{-} & \cite{tao2022seizure, lin2021driving, chang2021MMN-schizophrenia, li2022eeg, raeisi2022neonatal-epilepsy, asadzadeh2022bayes, li2021multidomain, song2021variationalinstance, du2022multi, xu2022dagam, zheng2021hiearachy, xue2022adaptive, zeng2020hierarchy, jia2022cr-gcn, tao2022seizure, li2021mutualgraphnet, bao2022linking, jia2022efficient, chen2022exploring, li2022attentionsleep, jia2021multi} \\ 
   \hline
\end{tabular}
\label{table: graph definition}
\end{table*}

\begin{table*}[ht]
\centering
\caption{Overview of methods in defining the input node features} 
\begin{tabular}{|m{0.15\textwidth}<{\centering}|m{0.1\textwidth}<{\centering}|m{0.1\textwidth}<{\centering}|m{0.1\textwidth}<{\centering}|m{0.4\textwidth}<{\centering}|}
  \hline
Method & Time domain & Frequency domain & Graph Domain & Papers \\ 
  \hline
Differential entropy & \ding{51} & \ding{55} & \ding{55} & \cite{zhong2020GNN-regularised-emotion, li2021self-organising-emotion, li2021self-organising-emotion, li2022eeg, yin2021LSTMfusion, chen2021bayesianGNN, li2021multidomain, jia2020graphsleepnet, wiercinski2022emotion, song2020instance, song2021variationalinstance, du2022multi, sun2022transformer, wang2021depression, zeng2022siam, zhang2019gcb, jang2018video, kong2022causalGNN, zheng2021hiearachy, xue2022adaptive, li2021mutualgraphnet, zhu2022locally, bao2022linking, li2021attentionLSTM} \\ 
  Raw signal & \ding{51} & \ding{55} & \ding{55} & \cite{demir2021ERP-comparison, kwak2020BCI-Fusion-GNN, chang2021MMN-schizophrenia, li2019GNN-motor-movement, asadzadeh2022bayes, zhdanov2022schizophrenia, sun2021adaptive, li2021multidomain, li2021dualGAT, hou2020biLSTM-GCN, hou2022gcns, li2021active, ding2021lggnet, priyasad2022affect, tao2022seizure, wang2022coteaching, dissanayake2021geometric, feng2022eeg, he2022spatial, chen2022exploring, li2022attentionsleep, lu2022pearnet, jia2021multi, shan2022STGCN-AD} \\ 
  Fourier Transform & \ding{55} & \ding{51} & \ding{55} & \cite{li2022attention-dynamic, tang2021selfsupervised, covert2019temporal} \\ 
  Power Spectral Density/Band Power & \ding{55} & \ding{51} & \ding{55} & \cite{raeisi2022neonatal-epilepsy, zhang2019gcb, jang2018video, kong2022causalGNN, zheng2021hiearachy, zeng2020hierarchy, jia2022cr-gcn, jia2022efficient, sun2022complex, klepl2022GNN} \\ 
  Graph theory metrics & \ding{55} & \ding{55} & \ding{51} & \cite{liu2022MST-GNN, ghosh2021decoding} \\ 
  Descriptive statistics & \ding{51} & \ding{55} & \ding{55} & \cite{lin2021driving, zeng2020hierarchy, jia2022efficient} \\ 
   \hline
\end{tabular}
\label{table: node features}
\end{table*}

The first part of the input to a GNN model is the brain graph structure inferred from the EEG data itself (Fig. \ref{fig:overview}A). We summarise the methods for defining the brain graphs in Table \ref{table: graph definition}. These methods can be generally categorised as learnable or pre-defined. 

An alternative categorisation of the brain graph structures is the functional (FC) and the ``structural" connectivity (SC). Generally, SC graphs are pre-defined, whereas FC graphs can be both pre-defined and learnable. SC in the classical sense of physical connections between brain regions is not possible to obtain using EEG signals since these are recorded at the scalp surface. Instead, we use the term to describe methods that construct brain graphs based on the physical distance between EEG electrodes. In contrast, FC refers to pairwise statistical relationships between EEG signals. 

SC graph is pre-defined such that electrodes are connected by an edge in the following way:
\begin{equation}
    e_{ij}= 
\begin{cases}
    1 \text{ or } 1/d_{ij},& \text{if } d_{ij} \leq t\\
    0,              & \text{otherwise}
\end{cases},
\end{equation}
where $e_{ij}$ is the edge weight connecting nodes $i$ and $j$, $d_{ij}$ is a measure of distance between EEG electrodes, and $t$ is a manually defined threshold controlling the graph sparsity.

Such an approach offers several advantages. First, the SC graph is insensitive to any noise effects of EEG recording since it is independent of the actual signals. Second, all data samples share an identical graph structure, provided the same EEG montage was utilised during the recording. This offers explainability advantages when combined with spectral GNN since the graph frequency components defined by the eigenvectors of graph Laplacian are fixed. On the other hand, the SC graph is limited to short-range relationships. Thus, it might not accurately represent the underlying brain network. Some papers propose to overcome this limitation by manually inserting global \cite{du2022multi,ding2021lggnet, zheng2021hiearachy, zeng2020hierarchy, chen2022exploring} or inter-hemispheric edges \cite{zhong2020GNN-regularised-emotion, li2022eeg, xu2022dagam}.

In contrast, an FC graph can be obtained from either classical FC measures (FC measure in Table \ref{table: graph definition} or learnable methods (e.g. feature concatenation/distance and attention methods in Table \ref{table: graph definition}). We refer to all of these methods as FC because they all measure the degree of interaction between two nodes, thus falling within the traditional definition of FC. Unlike SC, the FC graph is unique for each data sample and can contain both short- and long-range edges. On the other hand, since it is derived directly from EEG signals, it might be sensitive to noise. 

Learnable FC based on node feature distance or feature concatenation are generally computed as:
\begin{align}
   e_{ij} &= \theta_1(|h_i-h_j|) \text{ and} \\
   e_{ij} &= \theta_2(h_i \mathbin\Vert h_j),
\end{align}
respectively, where $\theta_1(\cdot)$ and $\theta_2(\cdot)$ are neural networks with input-output dimensions of $\mathbb{R}: d \rightarrow 1$ and $\mathbb{R}: 2\times d \rightarrow 1$, respectively; $|\cdot|$ denotes absolute value; $\mathbin\Vert$ denotes concatenation and $h_i$ is the node feature/embedding of node $i$. We discuss the attention-based graphs together with the types of graph convolutional layers in Section \ref{section: GNN type} and thus skip these methods in this section.

Special cases of brain graph definition are the shared-mask methods. These methods defined a matrix of learnable parameters with the same shape as the adjacency matrix of the input graphs that acts as a mask/filter by multiplying it with the adjacency matrix. This learnable matrix is a part of the model. Thus, the same mask is applied to all input graphs. However, a shared mask limits the size of the input graphs, i.e. the number of nodes must remain fixed so that the adjacency matrix can be multiplied with the shared mask.

In the current stage, which method should be preferred for brain graph classification tasks is unclear. Some authors attempt to avoid this issue by combining multiple methods. However, we instead suggest that the researchers carefully consider each of the presented methods in the context of the given classification task, as each method poses its unique set of strengths and weaknesses.

\section{Node Feature Definitions}

The second part of the input to a GNN model is the node feature matrix (Fig. \ref{fig:overview}A). We summarise the various definitions of node features in Table \ref{table: node features}. We categorise these definitions based on which domain they are computed, i.e. time, frequency and graph domains.

The time-domain methods are the most commonly used in the current literature. In particular, these are the differential entropy (DE) and raw signal methods. The popularity of DE is given by the fact that many of the open EEG datasets include this feature, such as the SEED \cite{zheng_investigating_2015} emotion recognition dataset. DE describes the complexity of a continuous variable and is defined as:
\begin{equation}
    DE(X) = -\int_{X} f(x)log(f(x)) \,dx \,
\end{equation}
where $X$ is a random continuous variable and $f(x)$ is the probability density function.

Many papers define the node feature as the raw EEG signal. However, the raw signal can be too long for a GNN to process effectively. Thus, it is often coupled with node feature pre-processing module and spatio-temporal GNNs (See \ref{section: node preprocessing} and \ref{section: GNN type} respectively) to either reduce the dimensionality or to extract the temporal patterns contained within the signal effectively. An alternative to the raw signal node feature is descriptive statistics, such as mean, median or standard deviation.

\begin{table}[ht]
\centering
\caption{Overview of node feature pre-processing before GNN layers.} 
\begin{tabular}{|m{0.15\textwidth}<{\centering}|m{0.08\textwidth}<{\centering}|m{0.2\textwidth}<{\centering}|}
  \hline
Method & Trained separately & Papers \\ 
  \hline
1D CNN &  & \cite{demir2021ERP-comparison, li2021self-organising-emotion, zhdanov2022schizophrenia, li2021dualGAT, du2022multi, priyasad2022affect, dissanayake2021geometric, lu2022pearnet} \\ 
  Feature-wise attention weighting & \ding{55} & \cite{ghosh2021decoding} \\ 
  bidirectional LSTM & \ding{51} & \cite{hou2020biLSTM-GCN} \\ 
  Temporal CNN & \ding{55} & \cite{li2021active, ding2021lggnet, lu2022pearnet} \\ 
  WaveletCNN & \ding{55} & \cite{li2021active} \\ 
  SincCNN & \ding{55} & \cite{priyasad2022affect} \\ 
  MLP & \ding{55} & \cite{sun2022complex} \\ 
  CNN Feature Extractor & \ding{51} & \cite{jia2021multi} \\ 
   \hline
\end{tabular}
\label{table: feature preprocessing}
\end{table}

Frequency-domain node features are usually defined as the Fourier frequency components obtained by the Fourier transform or the power spectral density. Both of these methods attempt to quantify the strength of various frequency components within the EEG signal. An advantage of these representations is their relatively low dimensionality compared to the raw signal described previously. 

Finally, graph-theoretical features can be utilised to describe the nodes, e.g. mean node weight \cite{liu2022MST-GNN} and betweenness centrality \cite{ghosh2021decoding, liu2022MST-GNN}. A severe limitation of this method is that the graph structure needs to be defined prior to node feature extraction. Thus, this node feature type is incompatible with learnable brain graph methods. 

\subsection{Node Feature Preprocessing}
\label{section: node preprocessing}

An optional next step after node features construction is some kind of node feature pre-processing module (NFP) (Fig. \ref{fig:overview}B). We summarise the types of NFPs in Table \ref{table: feature preprocessing}.

Most of the NFPs are integrated within the GNN architecture, thus allowing the model to be trained in an end-to-end manner. The exceptions are methods that utilise a pre-trained feature extraction neural network implemented as a bidirectional LSTM \cite{hou2020biLSTM-GCN} or a CNN \cite{jia2021multi}.

The surveyed NFPs are all based on a neural network. In most cases, these are variants of a CNN and multilayer perceptron (MLP). These modules aim to (1) reduce the dimensionality of the node features and (2) enhance the node features, including potentially suppressing noise or redundant information.

\section{Type of Graph Convolutional Layer}
\label{section: GNN type}

\begin{table*}[ht]
\centering
\caption{Overview of graph convolutional layers.} 
\begin{tabular}{|m{0.25\textwidth}<{\centering}|m{0.07\textwidth}<{\centering}|m{0.07\textwidth}<{\centering}|m{0.076\textwidth}<{\centering}|m{0.41\textwidth}<{\centering}|}
  \hline
Method & Spatial & Spectral & Temporal & Papers \\ 
  \hline
Graph Isomorphism Network & \ding{51} & \ding{55} & \ding{55} & \cite{liu2022MST-GNN, demir2021ERP-comparison, tao2022seizure} \\ 
  (Simplified) Graph Convolution Network & \ding{51} & \ding{55} & \ding{55} & \cite{zhong2020GNN-regularised-emotion, zhao2021seizure-focalloss, zhdanov2022schizophrenia, li2021multidomain, du2022multi, sun2022transformer, wang2021depression, xu2022dagam, ding2021lggnet, zeng2020hierarchy, feng2022eeg, jia2022efficient, klepl2022GNN} \\ 
  Chebyshev Graph Convolution & \ding{55} & \ding{51} & \ding{55} & \cite{kwak2020BCI-Fusion-GNN, chang2021MMN-schizophrenia, raeisi2022neonatal-epilepsy, asadzadeh2022bayes, yin2021LSTMfusion, chen2021bayesianGNN, khaleghi2023visualdecoding, zeng2022siam, hou2020biLSTM-GCN, zhang2019gcb, hou2022gcns, jang2018video, kong2022causalGNN, zheng2021hiearachy, jia2022cr-gcn, wang2022coteaching, dissanayake2021geometric, bao2022linking, li2021attentionLSTM, sun2022complex} \\ 
  Graph Attention Network & \ding{51} & \ding{55} & \ding{55} & \cite{li2022attention-dynamic, ghosh2021decoding, li2021dualGAT, priyasad2022affect, zhu2022locally, he2022spatial, chen2022exploring, lu2022pearnet} \\ 
  Diffusion recurrent gated & \ding{55} & \ding{51} & \ding{55} & \cite{tang2021selfsupervised} \\ 
  Spatio-temporal GNN (Spectral) & \ding{55} & \ding{51} & \ding{51} & \cite{sun2021adaptive, covert2019temporal, li2021mutualgraphnet, shan2022STGCN-AD, jia2020graphsleepnet, wiercinski2022emotion} \\ 
  Spatio-temporal GNN (Spatial) & \ding{51} & \ding{55} & \ding{51} & \cite{li2019GNN-motor-movement, li2022attentionsleep} \\ 
  Powers of Adjacency Matrix GNN & \ding{51} & \ding{55} & \ding{55} & \cite{song2020instance, song2021variationalinstance} \\ 
  GraphSAGE & \ding{51} & \ding{55} & \ding{55} & \cite{demir2021ERP-comparison, covert2019temporal} \\ 
  Spectral GNN & \ding{55} & \ding{51} & \ding{55} & \cite{li2021self-organising-emotion, li2022eeg} \\ 
  B-Spline Kernel GCN & \ding{51} & \ding{55} & \ding{55} & \cite{lin2021driving} \\ 
  Residual GCN & \ding{51} & \ding{55} & \ding{55} & \cite{li2021active} \\ 
  Multibranch architectures & \textbf{-} & \textbf{-} & \textbf{-} & \cite{li2021dualGAT, zeng2022siam, xue2022adaptive, zeng2020hierarchy, wang2022coteaching, li2022attentionsleep} \\ 
   \hline
\end{tabular}
\label{table: GNN type}
\end{table*}

A core part of a GNN model are the graph convolutional layers (GCN) (Fig. \ref{fig:overview}C). We summarise the utilised types of GCNs in Table \ref{table: GNN type}. We further categorise them based on the type of GNN as introduced in Section \ref{section: background}, i.e. spatial, spectral. Additionally, we add the temporal category, which is not a type of standalone GCN layer but must be combined with spatial or spectral GCN.

Interestingly, ChebConv is used in the majority of the surveyed papers (counting both ChebConv and spectral spatio-temporal GNN in Table \ref{table: GNN type}). Since EEG typically uses 128 electrodes in high-density montages, the size of the brain graphs is relatively small. In such cases, even a full spectral GNN would not be too computationally expensive for EEG classification. Therefore, it remains unclear why many authors opt for the ChebConv approximation of spectral GNN. We speculate that the influence of classical signal processing tools in EEG analysis might also serve as a sufficient argument for using spectral GNNs for EEG classification.

On the other hand, the other half of the surveyed papers experiment with a wide range of spatial GNNs. The (simplified) GCN is a popular method amongst these, which is equivalent to a 1st-order ChebConv ($K=1$). A special case of spatial GNN is the graph attention network (GAT). GAT allows for adjusting the graph by re-weighting the edges using an attention mechanism. Generally, the attention mechanism for computing the new softmax-normalised edge weight $e_{ij}$ is defined as follows:
\begin{equation}
e_{i,j} =
        \frac{
        \exp\left(\sigma\left(\mathbf{w}^{\top}
        [\mathbf{W}h_i \, \Vert \, \mathbf{W}h_j]
        \right)\right)}
        {\sum_{k \in \mathcal{N}(i)}
        \exp\left(\sigma\left(\mathbf{w}^{\top}
        [\mathbf{W}h_i \, \Vert \, \mathbf{W}h_k]
        \right)\right)},
\end{equation}
where $w$ and $W$ are the learnable parameters of the model, $\sigma$ is an activation function, $h$ is the node feature vector/embedding, and $N(i)$ is the set of nodes connected to node $i$. The resulting edge weights can then be passed to Equation \ref{eq: spatial GNN}.

Next, the spatio-temporal GNNs were tested for EEG classification in several instances. A spatio-temporal block consists of one GCN layer and one 1D-CNN applied temporally. This structure allows the model to extract both spatial (i.e. graph) and temporal patterns. There are both spatial and spectral variants of spatio-temporal GNN, and there is no indication as to which one should be preferred as no comparative study exists to date.

Finally, several papers adopt multi-branch architectures. These methods utilise multiple GCN layers applied in parallel to allow the model to focus on various aspects (also views) of the input graph. An example of such a model utilises two-branch GNN to learn from both FC- and SC-based brain graph structure \cite{li2022attentionsleep}. Alternatively, the individual frequency bands of EEG signals can be used to construct various graph views \cite{sun2022complex}.

\section{Node Pooling Mechanisms}

\begin{table}[ht]
\centering
\caption{Overview of node pooling mechanisms.} 
\begin{tabular}{|m{0.45\linewidth}<{\centering}|m{0.18\linewidth}<{\centering}|m{0.2\linewidth}<{\centering}|}
  \hline
Method & Learnable & Papers \\ 
  \hline
TopK & \ding{51} & \cite{chang2021MMN-schizophrenia, chen2022exploring} \\ 
  Hierarchical tree pooling & \ding{51} & \cite{liu2022MST-GNN} \\ 
  SortPool & \ding{51} & \cite{demir2021ERP-comparison} \\ 
  EdgePool & \ding{51} & \cite{demir2021ERP-comparison} \\ 
  SAGPool & \ding{51} & \cite{demir2021ERP-comparison, xu2022dagam} \\ 
  Set2Set & \ding{51} & \cite{demir2021ERP-comparison} \\ 
  Manual Clustering & \ding{55} & \cite{song2020instance, song2021variationalinstance} \\ 
  Graclus Clustering & \ding{55} & \cite{hou2022gcns, wang2022coteaching} \\ 
   \hline
\end{tabular}
\label{table: node pooling}
\end{table}

In some instances, reducing the number of nodes in the graph might be desirable. This can be achieved with a node pooling module (Fig. \ref{fig:overview}D). We summarise the node pooling modules utilised in the surveyed papers in Table \ref{table: node pooling}. 

There are both learnable and non-learnable node pooling modules in the literature. Please see the corresponding papers for a detailed description of these methods (Table \ref{table: node pooling}). Node pooling modules remain a relatively unexplored topic in the EEG-GNN classification models. Node pooling can (1) remove redundant nodes, (2) reduce the size of the graph embedding in a setting where the concatenation of node embeddings forms it, and (3) aid in the explainability of the model by identifying node importance with respect to the classification task.

\section{From node embeddings to graph embedding}

\begin{table*}[ht]
\centering
\caption{Overview of methods for the formation of graph embedding from a set of node embeddings} 
\begin{tabular}{|m{0.2\textwidth}<{\centering}|m{0.08\textwidth}<{\centering}|m{0.5\textwidth}<{\centering}|}
  \hline
Method & Learnable & Papers \\ 
  \hline
Sum readout & \ding{55} & \cite{liu2022MST-GNN, zhong2020GNN-regularised-emotion, bao2022linking} \\
  Average readout & \ding{55} & \cite{raeisi2022neonatal-epilepsy, li2021multidomain, xu2022dagam, covert2019temporal, jia2022efficient, chen2022exploring, sun2022complex} \\ 
  Maximum readout & \ding{55} & \cite{zhdanov2022schizophrenia, xu2022dagam, hou2020biLSTM-GCN, chen2022exploring, klepl2022GNN} \\ 
  Concatenate node embeddings & \ding{55} & \cite{lin2021driving, li2021self-organising-emotion, kwak2020BCI-Fusion-GNN, chang2021MMN-schizophrenia, li2022eeg, zhao2021seizure-focalloss, song2020instance, song2021variationalinstance, khaleghi2023visualdecoding, li2021dualGAT, zhang2019gcb, hou2022gcns, jang2018video, li2021active, ding2021lggnet, zheng2021hiearachy, xue2022adaptive, zeng2020hierarchy, jia2022cr-gcn, priyasad2022affect, wang2022coteaching, li2021mutualgraphnet, zhu2022locally, dissanayake2021geometric, jia2021multi, shan2022STGCN-AD} \\ 
  CNN-like Average/Maximum Pooling & \ding{55} & \cite{li2019GNN-motor-movement, feng2022eeg} \\ 
  SortPool & \ding{51} & \cite{bi2022multi} \\ 
  Attention weighted & \ding{51} & \cite{li2022attention-dynamic, zeng2022siam, li2022attentionsleep} \\ 
  CNN & \ding{51} & \cite{asadzadeh2022bayes, sun2021adaptive, kong2022causalGNN} \\ 
  LSTM & \ding{51} & \cite{yin2021LSTMfusion, li2021attentionLSTM} \\ 
  Capsule Network & \ding{51} & \cite{ghosh2021decoding} \\ 
  Transformer & \ding{51} & \cite{sun2022transformer} \\ 
  Bidirectional LSTM & \ding{51} & \cite{li2021dualGAT, feng2022eeg, he2022spatial} \\ 
   \hline
\end{tabular}
\label{table: graph embedding}
\end{table*}

The output of the graph convolutions is a set of learned node embeddings. Node embeddings in this form are suitable for tasks such as node classification and link prediction. However, for graph classification, the set of node features needs to be transformed into a unified graph representation (Fig. \ref{fig:overview}E). We summarise the methods for this transformation in Table \ref{table: graph embedding}.

The most straightforward method to form a graph embedding is to simply concatenate the node features. This approach poses a few limitations. First, the resulting graph embedding grows with the number of nodes, thus, the classification layer requires a large number of parameters. Second, all input graphs need to have the same number of nodes, limiting the model's generalisation to other datasets. Finally, such an approach is likely to include redundant or duplicated information in the graph embedding since GNN produces node embeddings by aggregating information from neighbouring nodes.

A readout function is one of the methods to form a graph embedding that addresses these issues. A readout forms the embedding by passing the node features through a permutation-invariant function. A general definition of a readout to obtain graph embedding of a graph $G_i$ from a set of $V$ node embeddings $H = [h_1,...,h_V]$ is given by:
\begin{equation}
    G_i = \sum^V_{k=1} h_k,
\end{equation}
where $\sum$ can be any permutation-invariant function. In the surveyed papers, these functions were sum, average and maximum. A few papers also experiment with attention-weighted sum to attenuate the role of unimportant nodes within the graph embedding \cite{li2022attention-dynamic}. An interesting alternative is to apply CNN-style average or maximum pooling node-wise \cite{li2019GNN-motor-movement}.

Alternatively, researchers explored various neural network models to obtain graph embeddings, such as CNN \cite{asadzadeh2022bayes, sun2021adaptive, kong2022causalGNN}, (bi-)LSTM \cite{yin2021LSTMfusion, li2021attentionLSTM, li2021dualGAT, feng2022eeg, he2022spatial}, Transformer \cite{sun2022transformer} and capsule networks \cite{ghosh2021decoding}. Additionally, graph pooling methods, such as DiffPool \cite{ying_hierarchical_2018}, SAGPool \cite{lee_self-attention_2019}, iPool \cite{gao_ipoolinformation-based_2022}, TAP \cite{gao_topology-aware_2021} and HierCorrPool \cite{wang_multivariate_2023} can be used for this purpose.

\section{Discussion}
Despite most of the surveyed papers being relatively recent, a wide range of GNN-based methods has already been proposed to classify EEG signals in a diverse set of tasks, such as emotion recognition, brain-computer interfaces, and psychological and neurodegenerative disorders and diseases (Fig \ref{fig: classification task}). This recent rise in popularity of GNN models for EEG might be attributed to (1) the development of new GNN methods and (2) advances in network neuroscience inspired an extension of this framework to deep learning. GNNs offer unique advantages over other deep learning methods. This is mainly the possibility of modelling multivariate time series and interactions among them with a single GNN model, which is not possible with CNN or recurrent networks. Additionally, patterns learned by GNNs can readily be interpreted in the context of network neuroscience, thus enabling a wide range of avenues for model explainability.

This survey categorises the proposed GNN models in terms of their inputs and modules. Specifically, these are brain graph structure, node features and their preprocessing, GCN layers, node pooling mechanisms, and formation of graph embeddings. This categorisation allows us to provide a quick and simple overview of the different methods presented in the EEG-GNN literature, appreciate the current state of the art in this field and identify promising future directions.

\subsection{Limitations of Surveyed Papers}
Surprisingly, we have identified the least variety and innovation in the category of GCN layers (Table \ref{table: GNN type}). A significant proportion of the surveyed papers utilise either ChebConv or ``vanilla" spatial GCN. This might be due to the relative novelty of the EEG-GNN field, and thus, many papers explore other areas of model design, such as node features and brain graph definitions. A few papers seem to successfully experiment with more complex types of GCN layers \cite{lin2021driving, li2021active,tang2021selfsupervised} and multi-branch architectures \cite{li2021dualGAT, zeng2022siam, xue2022adaptive, zeng2020hierarchy, wang2022coteaching, li2022attentionsleep}.

A major limitation of most surveyed papers is the lack of generalisability to external datasets that might use a different number of EEG signals. This is caused by (1) the use of ChebConv and (2) forming graph embedding by node feature concatenation \cite{lin2021driving, li2021self-organising-emotion, kwak2020BCI-Fusion-GNN, chang2021MMN-schizophrenia, li2022eeg, zhao2021seizure-focalloss, song2020instance, song2021variationalinstance, khaleghi2023visualdecoding, li2021dualGAT, zhang2019gcb, hou2022gcns, jang2018video, li2021active, ding2021lggnet, zheng2021hiearachy, xue2022adaptive, zeng2020hierarchy, jia2022cr-gcn, priyasad2022affect, wang2022coteaching, li2021mutualgraphnet, zhu2022locally, dissanayake2021geometric, jia2021multi, shan2022STGCN-AD}. (1) can be addressed by utilising spatial GCN layers as suggested above, and (2) can be solved by using a readout function or a suitable node pooling mechanism, which coarsens the graph to a fixed number of nodes. Additionally, there is a general lack of transfer learning experiments for EEG-GNN models, which might be a promising direction for future research.

Finally, we have identified an interesting gap in EEG-GNN research: the lack of utilising frequency band information in a more complex way. A few papers train separate models for each frequency band in isolation \cite{liu2022MST-GNN, zhong2020GNN-regularised-emotion, lin2021driving}. Alternatively, they propose concatenating the graph embeddings generated from the frequency-band-GNN branches \cite{li2022eeg, song2020instance, sun2021adaptive}.

\subsection{Future Directions}
Several promising directions can be identified in the rapidly evolving landscape of EEG-GNN research. First, a comprehensive comparison of the various GCN layers (e.g. spatial GNN, ChebConv, GAT and graph transformer) with respect to their influence on classification performance should be carried out to address this crucial design question in a systematic manner.

Second, enhancing the generalisability of models by addressing issues related to the varying number of EEG signals/electrodes and exploring transfer learning approaches can open new avenues for research. For instance, pre-trained GNN models on cheap-to-obtain large datasets, such as open databases for emotion recognition or BCI applications, would allow the application of complex GNN architectures to problems with limited data availability due to the high costs or small populations (e.g. clinical data, rare diseases and disorders). Focusing on these issues would likely improve the generalisability of the models when evaluated on a diverse set of EEG datasets and different classification tasks.

Lastly, the rich frequency information of EEG signals should be explored more. For instance, we suggest a plausible utility of integrating cross-frequency coupling (CFC) approaches into EEG-GNN models. There is growing evidence in the literature concerning the advanced brain functions (e.g. learning, memory) enabled by CFC \cite{jirsa_cross-frequency_2013}. Thus, integrating findings from neuroscience research into the EEG-GNN design promises both performance and explainability gains.

\subsection{Limitations of Our Survey}
It is worth noting that this paper does not follow a systematic review methodology; therefore, we do not assert that our findings are exhaustive. Instead, our objective is to offer a succinct and cohesive overview of the current research on EEG-GNN models to facilitate the development of innovative approaches and assist researchers new to this field.

One of the major parts of EEG-GNN models we omit in this survey is the model explainability. We suggest that a survey paper is not well suited for comprehensively covering this aspect of research. Instead, we suggest a comparative experimental study to be better suited to explore the various explainability options of GNN explainability. However, to maintain the comprehensiveness of this survey, we list the papers that report the use of certain methods of model explainability: \cite{li2019GNN-motor-movement, zhdanov2022schizophrenia, tang2021selfsupervised, sun2022transformer, jang2018video}.

\section{Conclusion}
In conclusion, this survey examined the current research on EEG-GNN models for classifying EEG signals. Various GNN-based methods have been proposed for tasks such as emotion recognition, brain-computer interfaces, and psychological and neurodegenerative disorders. The surveyed papers were categorised based on inputs and modules, including brain graph structure, node features, GCN layers, node pooling mechanisms, and graph embeddings.

GNNs offer a unique method for analysing and classifying EEG in the graph domain, thus allowing the exploitation of complex spatial information in brain networks that other neural networks do not. Additionally, GNNs can be easily extended with CNN and recurrent network-based modules at various stages of the GNN architecture, such as for node feature pre-processing, node embedding post-processing and graph embedding formation.

However, limitations and areas for improvement were identified. There is a lack of variety and innovation in GCN layers, with many papers utilising ChebConv or ``simple'' spatial GCN without clear justification. Generalisability to external datasets with varying numbers of EEG electrodes is limited. Transfer learning experiments and integration of cross-frequency coupling approaches are potential future research to enhance the performance and explainability of GNN.

\newpage
\bibliographystyle{unsrt}
\bibliography{ref}

\end{document}